\documentclass[12pt]{article}
\usepackage{geometry}
\usepackage{hyperref}                          

\usepackage{graphicx}	
\usepackage{amsmath}	
\usepackage{amssymb}	
\usepackage{framed,color}

\usepackage{sidecap}
\usepackage{natbib}

\newcommand{\aap}{A\&A}
\newcommand{\aaps}{A\&AS}
\newcommand{\aj}{AJ}
\newcommand{\apj}{ApJ}
\newcommand{\apjl}{ApJ}
\newcommand{\apjs}{ApJS}
\newcommand{\araa}{ARA\&A}
\newcommand{\mnras}{MNRAS}

\begin{document}

\justify

\vspace{5cm}
\textbf{
MASSIVE STARS IN THE THIRTIES: AWAITING NEW HST DISCOVERIES  }  

\begin{center}

{\bf The organizing committee of the IAU G2 Commission ``Massive Stars''}\\
\vspace{0.3cm}
{\bf Lidia M. Oskinova,
Sylvia Ekstr\"om, 
Miriam Garcia,
Takashi Moriya,\\
Andreas A.C. Sander, 
Sergio Sim\'on-D\'{\i}az, and
Aida Wofford}

\end{center}

\begin{figure}[h]
\centering
   \includegraphics[width=0.6\textwidth]{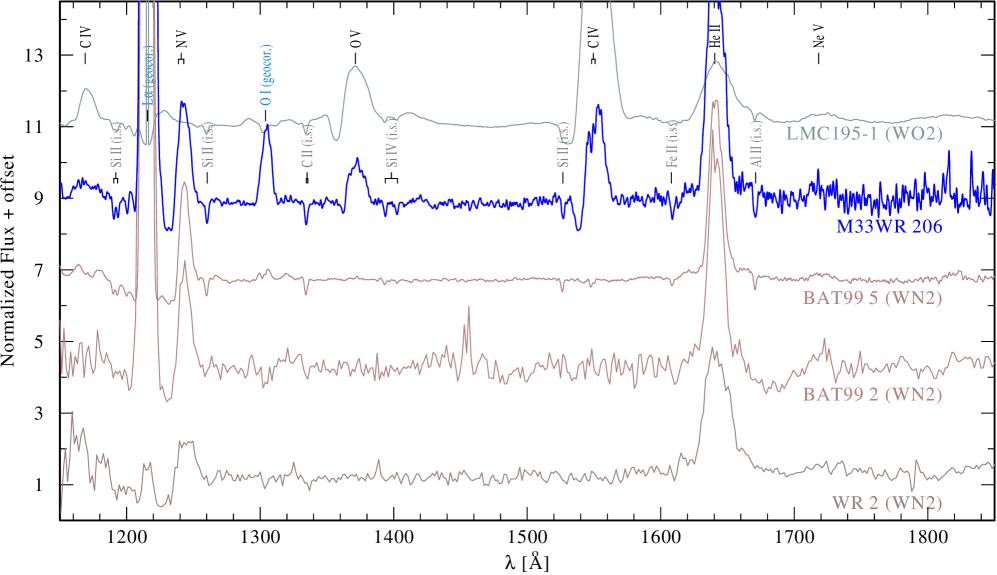}
   \caption{\footnotesize HST COS spectra of the hottest WR type stars of WN2/WO1, WN2, and WO2 subtypes. The spectra have been shifted along the ordinate for clarity. The WN2--WO2 stars emit a high flux of photons capable of fully ionizing helium. These hot stars are difficult to detect in integrated stellar populations, but at the same time they yield a hard ionizing flux. The figure is adopted from \citet{2026NatAs..10..290S}.}
   \label{fig:wruv}
\end{figure}

\noindent
Stars born with masses $>$$10\,M_\odot$ are the principle sources of stellar feedback. During large parts of  their lives, massive star spectral energy distribution peaks in the UV. They settle on the main sequence with OB-type  and  evolve to become cooler yellow and red supergiants (YSG, RSG). Some massive stars become blue supergiants (BSG) and may pass through a luminous blue variable (LBV) phase. Highly evolved hot hydrogen-depleted stars exhibit Wolf–Rayet (WR) spectral types (Fig.\,\ref{fig:wruv}). Massive stars drive winds that become stronger as the stars evolve. Binary interactions further increase the complexity of massive-star evolutionary pathways \citep{2007ARA&A..45..177C,2022ARA&A..60..455E,2022ARA&A..60..203V,2024ARA&A..62...21M}.

Revolutionary developments in astrophysics call for a much deeper understanding of massive stars and their final evolutionary products -- neutron stars and black holes.
Among the most prominent examples is the discovery of gravitational waves from merging black holes in 2016 \citep{2023MNRAS.525.2891C}. More recently, JWST revealed the prevalence of UV-luminous galaxies in the young Universe dominated by massive star clusters \citep{2025A&A...693A.271S,2026arXiv260421493M}. From a theoretical perspective, stellar feedback is firmly established as an essential ingredient in cosmological simulations \citep{2026MNRAS.548ag375S}. On smaller scales, the influence of massive stars and their explosive deaths is a key input in star-formation models, as well as in models of cosmic-ray production and particle acceleration in galaxies \citep{2018ApJ...857...57N,2025A&A...695A.175M}.

The study of massive stars and their feedback lies at the crossroads of modern astrophysics demanding  a more quantitative understanding which will continue to grow in the 2030s.

The HST is the workhorse of massive-star astrophysics.  In the 2030s, COS and STIS  will remain the most powerful UV spectrographs in terms of spectral resolution and sensitivity available to the broad international community. Among recent HST legacy programs  is the \href{https://ullyses.stsci.edu/}{ULLYSES}
 project dedicated  to creating a UV spectral library of massive stars in low-metallicity ($Z$) galaxies, demonstrating the strong value of massive star spectroscopy for astrophysics in general, as well as a number of large programs devoted to studies of
massive stars and clusters \citep{2015AJ....149...51C,2022ApJS..261...31B,2026ApJ..1000..241S}. Driven by the science, the demand for the HST  will increase in the next decade.

\section{Key questions which require HST's capabilities}


Massive star astrophysics is rapidly developing. Currently, three non-LTE stellar atmosphere codes are capable of producing high-fidelity synthetic spectra of hot stars and their stellar winds \citep{2024A&A...689A..30S}. HST data analyzed with the help of modern stellar atmosphere models provide results that rapidly propagate into stellar evolution and population synthesis models. Scientific progress in this field is advancing rapidly, and the answer to many fundamental questions over the next decade is strongly based on the capabilities of HST.

\begin{SCfigure}[]
\centering
   \includegraphics[width=0.55\textwidth]{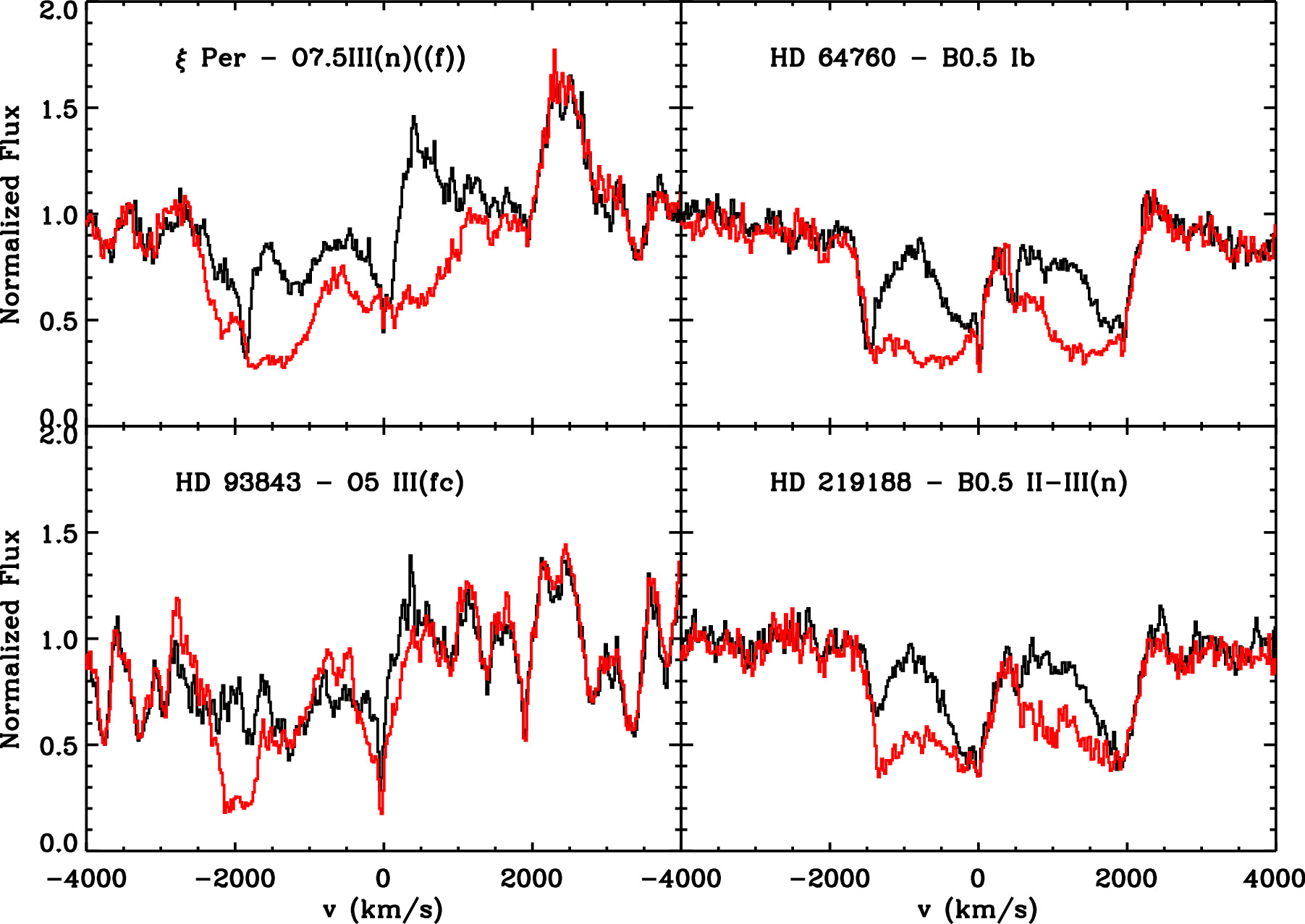}
   \caption{\footnotesize Spectra of Si\,{\sc iv} $\lambda1400$\AA\ doublet measured at two different epochs in
   four stars with a range of spectral types, demonstrating the ubiquity of spectral variability in the wind diagnostic 
   lines. Fitting spectral model to a single epoch observations results in 50--100\%\ uncertainty on the derived wind parameters (mass-loss rate, clumping, wind velocity field). The figure is adopted from \citet{2024ApJ...971..166M}.}
   \label{fig:uvv}
\end{SCfigure}

\smallskip
\noindent
{\color{blue} \it $\bullet$ {\em How does wind mass-loss rate scale with metallicity for stars of different masses and evolutionary stages?}  }
    Hot luminous stars drive stellar winds that remove mass and angular momentum. A long-standing problem is establishing the relationship between stellar wind strengths and the fundamental stellar parameters (mass, luminosity, temperature, and abundances -- baseline galactic and stellar).  Spectroscopic modeling of UV spectra is the most reliable 
    tool to measure stellar wind parameters. Currently, theory and models have been mainly tested and adjusted 
    by observations of  BSGs in the Galaxy and Magellanic Clouds \citep{2024A&A...692A..91V, 2024A&A...692A..88B}. However, there are apparent large discrepancies  between  theoretical expectations and observed wind signatures of the  main sequence OB stars and the products of binary interactions \citep{2022A&A...666A.189R, 2024ApJ...974...85T}. Furthermore, developments in recent years have shown that spectral disentanglement is necessary for correct interpretation of binary spectra, which requires a series of phase resolved spectra \citep{2023A&A...674A..56R}.  

    To address this challenge, time-series spectroscopy of representative samples of stars in galaxies spanning a range of metallicities is necessary.

\smallskip
\noindent 
{\color{blue} \it $\bullet$ {\em What time series of UV spectra reveal about the engines that drive stellar winds?}}
In 1990s, the International Ultraviolet Explorer (IUE) monitored UV spectra of ten  Galactic O stars  \citep{1996A&AS..116..257K}. This led to the unexpected discovery of P Cygni type line profiles exhibiting a  pattern of variability in the form of discrete absorption components  migrating through the absorption troughs. This pattern is different for each star but remains relatively constant for a given star. The subsequent hydrodynamic models explained the observed UV spectral variability by dynamics of winds in rotating stars which have spots
as was later confirmed by precise photometric observations \citep{1996ApJ...462..469C,2014MNRAS.441..910R}. By now it is clear that UV spectra of all OB-type stars are variable which strongly complicates measurements of stellar wind (Fig.\,\ref{fig:uvv} \citep{2024ApJ...971..166M}).  The resulting errors propagate to the stellar wind theory, stellar evolution and population synthesis models. 

HST spectroscopy has confirmed the complex structure of wind diagnostic lines in snapshot observations, but so far only limited efforts have been devoted to spectral monitoring \citep{2019ApJ...873...81M}. Theory suggests that the UV variability pattern should depend on metallicity, and some observational evidence for this has already been found in archival HST data \citep{2011A&A...534A.140C,2024MNRAS.52711422P}. However, key questions regarding the physical mechanisms, coupling stellar interiors and stellar winds, remain to be addressed in the future.

To address this challenge, high-resolution spectroscopic monitoring with a cadence shorter than the stellar rotation period is required for  stars in galaxies spanning a range of metallicities.

\smallskip
\noindent 
{\color{blue} \it $\bullet$ {\em How massive stars evolve, interact, and die?}}
Work on stellar evolution and population synthesis models will further intensify in the next  decade \citep{2024ARA&A..62...21M}. It is highly important to determine the key parameters governing these models, such as mixing, mass loss via stellar winds, and, in binaries,  stripping,  mass-transfer efficiencies and common-envelope parameterizations. To achieve new science, the predictions of evolutionary models must be verified observationally, and this can currently only be done with HST. Recent examples include attempts to identify and characterize stars stripped of their outer hydrogen-rich envelopes in binary systems. HST observations play a key role in identifying and characterizing such objects and provide support to the theoretical predictions on the mass exchange in binaries at different metallicities  \citep{2022A&A...662A..56K,2022A&A...659A...9P,2023Sci...382.1287D,2026A&A...708A.187M,2023ApJ...959..125G,2024A&A...692A..90R}.

\begin{SCfigure}[]
\centering
   \includegraphics[width=0.5\textwidth]{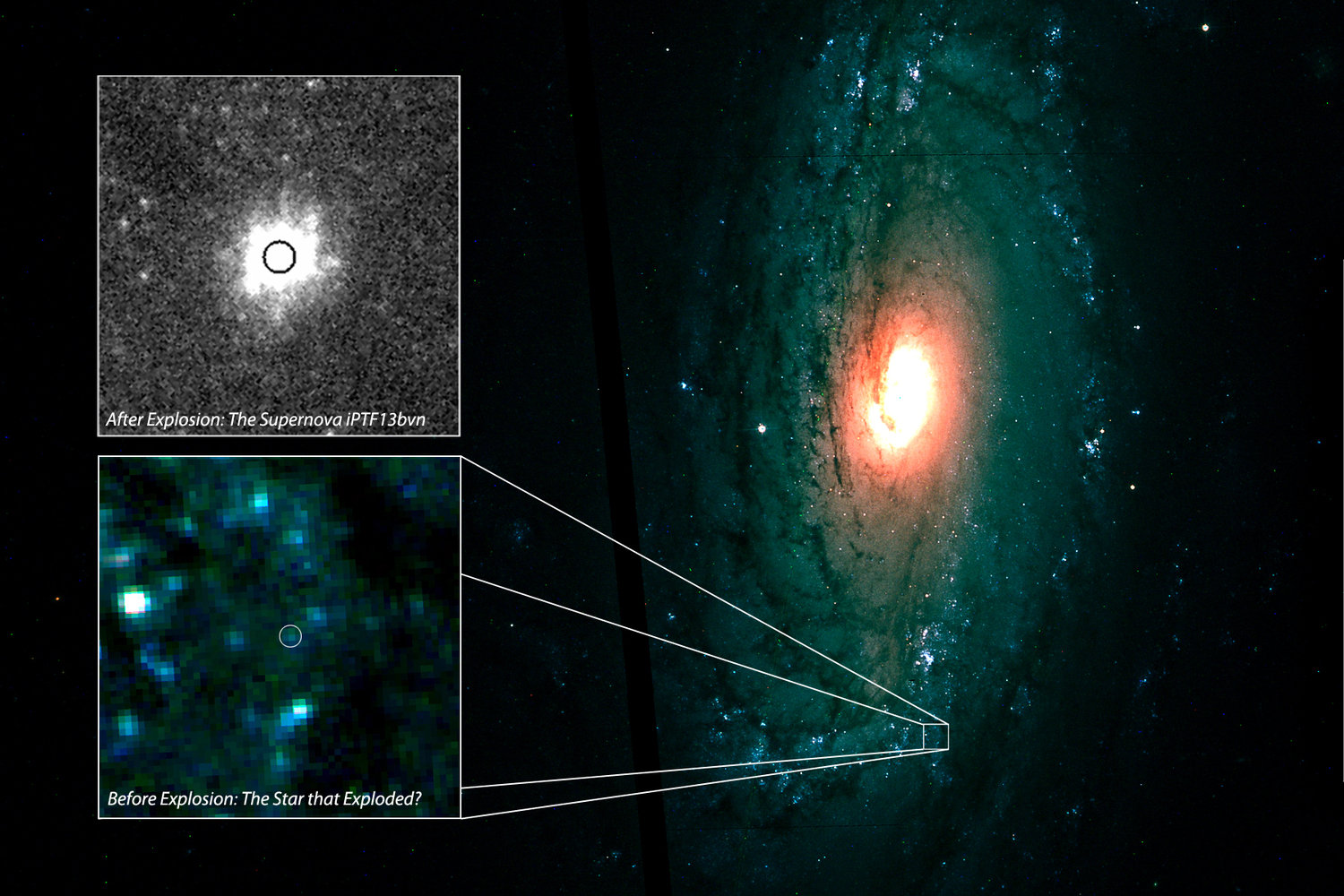}
   \caption{\footnotesize Using data from the Keck II and HST  (ACS F435W, F555W, and F814W)  , this composite image shows the supernova iPTF13bvn, and its possible progenitor WR-type star, in the galaxy NGC 5806. The SN iPTF13bvn was accompanied by a $\gamma$-ray burst \citep{2013ApJ...775L...7C,2013ApJ...776L..34S}. IPAC/JPL/Caltech}
   \label{fig:sn}
\end{SCfigure}

Another active area of research is related to phenomena accompanying the late evolutionary stages of massive stars and their core collapse. Advances in supernova explosion modeling provide guidance for the diverse types of transients that can accompany the final stages of stellar evolution. LBV-like eruptions and similar transient events associated with massive star evolution are often referred to as SN impostors \citep{2012ApJ...758..142K,2021ApJ...917...63A}. In particular, the question of mass ejections prior to core collapse has recently received considerable attention \citep{2026MNRAS.548f2261K}. HST imaging, photometry, and spectroscopic follow-up observations of core-collapse SNe are extremely valuable for constraining the final evolutionary stages of massive progenitor stars (Fig.\,\ref{fig:sn}).

Re-imaging archival fields in star-forming galaxies will provide a baseline of $\sim$$50$\,yr for transient science and  strongly contribute to the scientific legacy of HST \citep{2024arXiv240512297J}.

\smallskip
\noindent
{\color{blue} \it $\bullet$ {\em What are the largest stellar masses and how these depend on metallicity?}}
The upper cutoff and shape of the upper initial mass function (IMF) remain highly uncertain. However, establishing the masses of the most massive stars is of huge astrophysical significance, particularly in light of recent gravitational-wave discoveries and observations from JWST \citep{2024MNRAS.529.3301T,2025arXiv250818083T}. The current record holders are the $> 150\,M_\odot$ stars in the R\,136 cluster at the center of the 30\,Doradus star-forming region \citep{2010MNRAS.408..731C}. There are also indications of a top-heavy IMF in this region \citep{2018Sci...359...69S}.

\begin{SCfigure}[]
\centering
   \includegraphics[width=0.4\textwidth]{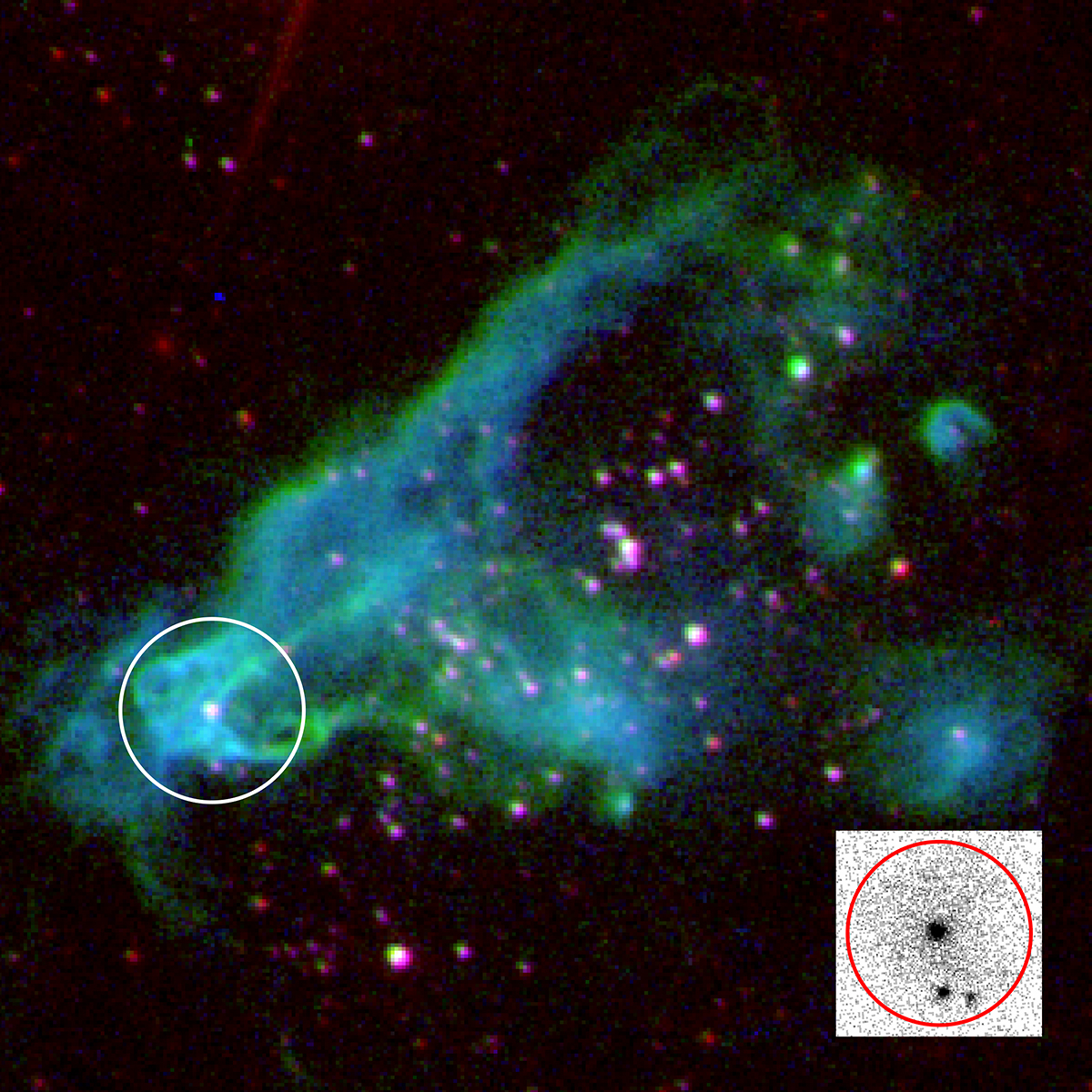}
   \caption{\footnotesize 
   False-color HST ACS image (502N blue, F658N green, and F814W red) of the area around the ULX Ho II X-1 with the adjacent H\,{\sc ii} region known as the ``Foot Nebula''.  The white circle highlights the He\,{\sc iii}bubble generated by ionising X-ray radiation emitted by the ULX. The ULX optical counterpart is seen as the white spot centered in the cycle. In the lower right corner, the acquisition image of the HST COS observations is displayed. The $2.5"$ COS diameter aperture (shown as a red circle) encompasses the ULX, the He\,{\sc iii} bubble, and two neighbouring UV-bright stars. The figure is adopted from \citet{2024A&A...690A.347R}.}
   \label{fig:hoii}
\end{SCfigure}

Empirically, the most promising sites in which to search for very massive stars are the cores of rich, massive star clusters. Resolving the most UV-luminous stars in these crowded environments and constraining their masses is a challenging task that can only be achieved with the help of HST. The first step toward measuring stellar masses is determining stellar luminosities, for which reliable photometric measurements from the UV to the IR are required. Spectroscopic analyses in the UV and optical, together with time-series observations, are then needed to further refine mass estimates.

To address stellar populations at the upper end of the IMF, targeted observations of massive star-forming regions preselected from previous HST legacy programs are required.

\smallskip
\noindent 
{\color{blue} $\bullet$ {\em\it{What are the properties of massive binary systems hosting compact objects?}}}
A natural outcome of massive binary evolution is systems containing  a compact object -- a white dwarf, a black hole, or, most commonly, a neutron star. The compact object may accrete matter from its massive-star companion; in this case, it becomes a bright X-ray source known as a high-mass X-ray binary (HMXB). The most luminous among these systems are ultra-luminous X-ray sources (see example in Fig.\,\ref{fig:hoii}). There are also binaries in which the compact object is X-ray faint. In such cases, only phase-resolved spectroscopy can reveal its presence \citep{2022NatAs...6.1085S,2022A&A...664A.159M}. In either case, HST observations are indispensable for identifying the optical/UV donor star counterparts of X-ray sources and determining its stellar and wind  properties  through photometry and spectroscopy. Only in this way we can obtain reliable mass measurements of compact objects, and constrain evolutionary scenarios as well as accretion regimes  \citep{2016ApJ...831..117F,2021ApJ...910...74B,2025A&A...698A..37R}.

In the 2030s, HST is needed to address this problem through dedicated observations of binary systems preselected from X-ray surveys, as well as of suspected systems containing compact objects that are quiescent in X-rays.

\section{Instrument capabilities and operational requirements}

While some planned missions provide partial UV spectral coverage tuned for specific science cases, the HST enables medium and high-resolution UV spectroscopy over a wide  wavelength range. E.g.\  since the end of the FUSE about 20\,yr ago, the range between $900$ and $1150$\,\AA\ can only be accessed with COS. Massive stars show important diagnostics only in this regime, which will be used to uncover very hot objects (or companions) via their O\,{\sc vi} and S\,{\sc vi} lines.

It is of paramount importance to maintain  all capabilities of HST in the UV. 
Imaging in optical and IR also remains important, in particular for transient science. 
Cross-calibration between HST and JWST is necessary for science which requires  long time base.

\smallskip\noindent
{\bf$\bullet$ ACS $\bullet$} The SBS FUV capabilities will remain uniquely suited for massive star 
studies in 2030s. Specifically, searching for and characterizing  elusive hot evolved descendants of primaries in 
binaries is one of the key questions in stellar astrophysics.

\smallskip\noindent
{\bf$\bullet$ WFC3 $\bullet$} Among UVIS filters, the  narrow band filters such as F469N, F680N, F658N, etc., are
indispensable for studies of massive star feedback, and in linking sub-pc scales of individual stars and clusters with the properties of the ionized ISM on $\propto 100$\,pc scale. Identifying individual extragalactic OB binaries with YSG and RSG yellow companions requires deep HST UV photometry, including $U$-band and FUV imaging with WFC3 and ACS, and followup spectroscopy with COS and STIS.

\smallskip\noindent
{\bf$\bullet$ COS $\bullet$} Both TIME-TAG and ACCUM modes are needed. For studies of stellar wind dynamics and precise mass-loss rate measurements, the high-resolving power $R>10000$ (G130M, G160M) 
corresponding to resolution in stellar wind velocity $v< 150$\,km\,s$^{-1}$ is required. G140L grating is excellent 
for spectroscopy of stars in galaxies at $\sim$\,Mpc distances $R>2000$ is necessary to detect the presence of stellar winds and estimate wind parameters.

\smallskip\noindent
{\bf$\bullet$ STIS $\bullet$}  
High resolution ($R> 30\,000$) UV echelle spectroscopy is the only possibility to obtain detailed information about stellar wind dynamics and measure abundances with high precision. New science will be enabled by searching for magnetic fields via Zeeman splitting of lines in UV/optical spectra \citep{2023Sci...381..761S}. HST's long-slit and slitless spectroscopy  is required to gain access to individual stars in extragakactic star-forming regions (Fig.\,\ref{fig:stis}).

\smallskip\noindent


\section{Towards the development of the HWO}
\label{sec:hwo}

\begin{SCfigure}[]
\centering
   \includegraphics[width=0.5\textwidth]{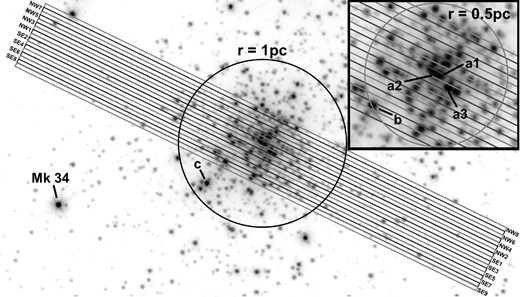}
   \caption{\footnotesize 
HST/STIS slits ($52"\times 0.2"$) superimposed upon an F336W WFC3/UVIS image of R136, together with a circle of radius $4.1"$ (1\,pc)  and identification of the acquisition star Melnick \,34. The active slit length for MAMA observations is the central 25 arcsec. The zoom highlights the central region, including identification of individual slits and the integrated R136a cluster.
The figure is adopted from \citet{2016MNRAS.458..624C}.}
   \label{fig:stis}
\end{SCfigure}

The HST data provide an essential anchor  for science cases that require long-baseline monitoring \citep{2024arXiv240512297J}.
For studies of massive star populations, an integral-field or multi-object UV spectrograph is essential \citep{2026ASPC..542....3S,2026ASPC..542...37M}. The long-slit STIS mosaics provide key scientific and technical input for defining  HWO instrumental requirements 
(Fig.\,\ref{fig:stis}). Importantly, continuation of HST imaging observations enables the long-term science, and the identifications of progenitors for the HWO transient science.

\section{Suggested community driven  Hubble Legacy Programs}

\smallskip\noindent
1. Re-imaging of fields in star-forming galaxies taken in the UV filters to identify blue transients and test the scenarios of silent collapse of hot luminous stars \citep{2026arXiv260412868G}.

\smallskip\noindent
2. High-cadence STIS monitoring of a selected sample of stars in the Galaxy, LMC, and SMC to establish patterns of spectral line variability and probe stellar activity cycles.

\smallskip\noindent
3. A systematic search for hot companions in binary stars and post binary-interaction products in the Galaxy, LMC, and SMC by using the UV photometry of suitably selected clusters. A follow-up spectroscopy of the well characterized sample of candiate stripped stars  will allow to draw robust conclusions on evolutionary models.

\pagebreak


\end{document}